\documentstyle[pre,aps]{revtex}        %%single-sided
\input epsf
\tighten
\begin{document}
%\draft
%\bibliographystyle{prsty}

\twocolumn[\hsize\textwidth\columnwidth\hsize\csname @twocolumnfalse\endcsname
\title{Finite driving rates in interface models of Barkhausen noise}
\author{
S. L. A. de Queiroz\footnote{Electronic address:sldq@if.ufrj.br} and
M. Bahiana\footnote{Electronic address:monica@if.ufrj.br}
} 
\address{
Instituto de F\'\i sica, Universidade Federal do Rio de Janeiro,\\
Caixa Postal 68528, 21945-970  Rio de Janeiro RJ, Brazil}
\date{\today}
\maketitle
\begin{abstract}
We consider a single-interface model for the description of Barkhausen
noise in soft ferromagnetic materials. Previously, the model had been
used only in the adiabatic regime of infinitely slow field ramping. We
introduce finite driving rates and analyze the scaling of event sizes and 
durations for different regimes of the driving rate. 
Coexistence of intermittency,
with non-trivial scaling laws, and finite-velocity interface motion 
is observed for high enough driving rates. 
Power spectra show a decay $\sim \omega^{-t}$, with $t<2$
for finite driving rates, revealing the influence of the internal 
structure of avalanches.
\end{abstract}
\pacs{PACS numbers:05.40.-a, 75.60.Ej, 05.65.+b, 75.50.Lk, 75.40.Mg}
\twocolumn
\narrowtext
\vskip0.5pc]
%\newpage

\section{Introduction} 
\label{intro} 
The Barkhausen effect~\cite{barkorig} constitutes a useful,
non-destructive probe into the
domain structure of soft ferromagnetic materials. By ramping an
externally applied magnetic field within appropriately chosen ranges of
intensity and driving rate, one causes microscopic domain wall
motions, i.e. magnetization jumps, within a sample. The consequent changes
in magnetic flux, in
turn, induce a time-dependent electromotive force $V(t)$ on a coil
wrapped around the sample. Analysis of $V(t)$, assisted by suitable
theoretical
modeling, may provide insight both into the domain structure itself and
its dynamical behavior.\par
Recent increased interest
in the problem stems partly from the successful application of methods
of non-equilibrium statistical mechanics, such as a Langevin
description via Fokker-Planck
equations~\cite{abbm} and avalanche 
models~\cite{umm,seth,zap,us,bt99,pds99,bt00,btn00,ks00,zcds,dz99,dz00}.
Though there is general agreement on the basic underlying mechanisms
of Barkhausen noise, pinning down specific details to features of
theoretical models has proved a complex task. For instance, while
several formulations have focused on the motion of a single interface
in a disordered medium~\cite{abbm,umm,zap,us}, others
have adopted a picture of nucleation of multiple domains in a random-field
Ising system~\cite{seth}. 

The rapid and discrete variation of the magnetization observed in experiments
is a direct manifestation of the existence of different time scales. Often 
the magnetization jumps are regarded as instantaneous events, but this 
is a simplification
that allows the description of a limited dynamical regime, 
as finite driving rates
are known to affect several aspects of the hysteresis cycle.

Barkhausen noise is observed in the central part of the hysteresis
cycle, near the coercive field where the magnetization process is
mainly due to domain wall motion, so interface models are usually
successful in describing the associated non-trivial 
scaling laws~\cite{umm,zap,us,zcds,dz00}.
In this paper we study the evolution of a single two-dimensional interface
as it advances driven at a finite rate. The scaling of event sizes and
durations is analyzed for different regimes of the driving rate.

\section{Model; finite driving rates} 
\label{intro2} 
Here we shall use the single-interface model introduced in
Ref.~\onlinecite{umm} for the description of Barkhausen noise, with
adaptations for a finite driving rate.
The interface at time $t$ is described,
in space dimensionality $d$, by its height 
$h(\vec{\rho_i},t)$, where
$\vec{\rho_i}$ is the position-vector of 
site $i$  in a $(d-1)$-dimensional lattice. Here we consider only $d=3$. 
At each $t$, the height function $h_i = h(\vec{\rho_i},t)$ is assumed to be 
single-valued, thus  the 
interface element corresponding to the $d$-dimensional position-vector 
$\vec{r_i}=(\vec{\rho_i},h_i)$ may be unambiguously labeled by $i$. 
Simulations are performed on an $L^{d-1} \times \infty$  geometry,
with the interface motion set along the infinite direction.
Therefore finite-size effects are controlled by the length parameter 
$L$~\cite{us}. 
Each element $i$ of the interface experiences a force of the form:
\begin{equation}
f_i=u(\vec{r_i})+\frac{k}{\gamma}\left[\sum_{j=1}^\gamma h_{\ell_j(i)}-
\gamma h_i\right]+H_e~,
\label{force}
\end{equation}
where 
\begin{equation}
H_e=H-\eta M~.
\label{He}
\end{equation}
The first term on the RHS of (\ref{force}) represents the pinning force,
$u$, and brings quenched disorder into the model by being chosen 
randomly, for each lattice site $\vec{r_i}$,  from a
Gaussian distribution of zero mean and standard deviation $R$. Large negative
values of $u$ lead to local elements where the interface will tend to be
pinned, as described in the simulation procedure below.
The second term corresponds to a cooperative interaction among interface 
elements,  assumed here to be of elastic (surface tension)
type. In this term, $\ell_j(i)$  is the position of the $j$-th nearest 
neighbor of site $i$ and $\gamma$ is the coordination number of the  
$(d-1)$-dimensional lattice  over which the interface projects.
The tendency of this term is to minimize 
height differences among interface sites: higher (lower) interface elements 
experience a negative (positive) force from their neighboring elements. 
The force constant $k$ gives the intensity of the elastic coupling, 
and is taken here as the unit for $f$.
The last term is the effective driving force, resulting from the applied 
uniform external field $H$ and a demagnetizing field which is taken to be 
proportional to 
$M=(1/L^{d-1})\sum^{L^{d-1}}_{i=1} h_i$, 
the magnetization  (per site) of the previously flipped spins for a
lattice of width $L$.

%{\bf Referee's point (1c)}

For actual magnetic samples, the demagnetizing field is not necessarily
uniform along the sample, as implied in the above expression; even when
it is (e.g. for a uniformly magnetized ellipsoid), $\eta$ 
would depend on the system's aspect ratio. Therefore, our approach amounts
to a simplification, which is nevertheless expected to capture the
essential aspects of the problem. See Ref.~\onlinecite{zcds} for a
detailed discussion.

%{\bf *****************************}

Here we use $R=5.0$, $k=1$, $\eta=0.05$, the same values as in the $d=3$
simulations of Ref.~\onlinecite{us}.

We start the simulation with a ~flat wall. All spins above it are
unflipped.
The applied field $H$ is set to
the saturation value of the effective field $H_e$, in order to minimize
transient effects (see e.g. Figure 1 of Ref.~\onlinecite{us}). The
saturation $H_e$ depends on $R$, $k$ and $\eta$ ({\em not} noticeably on
$L$)~\cite{us}, and can
be found from small-lattice simulations. The force $f_i$ is then
calculated for each unflipped site along the interface, and each spin at a
site with
$f_i\geq 0$ flips, causing the interface to  move up one step.
The magnetization is updated, and this process continues, with as many
sweeps of the whole lattice as necessary, until
$f_i<0$ for all sites, when the interface comes to a halt.
The external field is then increased by the minimum amount needed to bring
the most weakly pinned  element to motion. The  avalanche size corresponds 
to the number of spins flipped between two interface stops.

In line with standard practice~\cite{bt99,pds99,bt00,btn00,ks00} our basic
time unit is one lattice sweep, during which the external field is kept
constant, and all spins on the interface are probed sequentially as
described above.  In the adiabatic regime, the external field
is kept constant for the whole duration of an avalanche, i.e. for as many 
sweeps as it takes until 
no unstable sites are found along the interface. It is then increased by
the amount needed to flip the weakest one. At finite driving rates, the
field is increased by a fixed amount, henceforth denoted $\Delta$, at the
start of each sweep
while an avalanche is taking place. Eventually, no more unstable sites
will be left, and then one proceeds as in the adiabatic regime, increasing
the field as much as necessary to start a new avalanche. Calling $\delta
H$ this latter quantity, the time interval between the end of one event
and the start of the next is then $\delta H /\Delta$.    

In what follows, we usually collected samples of 10,000 avalanches for
each simulation, and data analysis has always been performed with the
whole sets of data; however, in figures such as scatter plots of duration
versus size we display only a representative subset, typically
%1,500--3,000 events, in order to avoid unnecessary clutter.  
500--1,000 events, in order to avoid unnecessary clutter.  

\section{Size and duration distributions} 
\label{std} 
We consider the single-interface model introduced in
Ref.~\onlinecite{umm}, initially in the adiabatic limit.
In Ref.~\onlinecite{us} we showed that, in the context of this model, the
upper cutoff in the avalanche size distribution is simply a manifestation
of finite-size effects, i.e. of the width $L$ of the cross-section of
our simulated systems perpendicular to the direction of interface advance.
Fitting our data to the customary power-law form multiplied by an 
exponential decay, namely
\begin{equation}
P(s) \propto s^{-\tau}\exp\left(-s/s_0\right)\ \ ,
\label{eq:psize}
\end{equation}
we had $\tau$ in the range $1.2-1.4$ and $s_0 \propto L^u$, $u=1.4 \pm
0.1$. \par
%{\bf Referee's point (1a)}

The above value for $u$ is consistent with $1/\sigma_k \simeq 2/3$ 
[ defined via $s_0 \sim k^{-1/\sigma_k}$, $k$ being a generic
demagnetization factor], found both from
renormalization-group analysis and numerical simulations of the
interface model of Refs.~\onlinecite{zap,zcds,dz00}, provided one
makes the identification $k \equiv 1/L^2$,
as pointed out in Ref.~\onlinecite{dz00}.

%{\bf *****************************}

We first investigate whether finite widths affect
properties related to avalanche duration. 
\begin{figure}
\epsfxsize=8.4cm
\begin{center}
\leavevmode
%\epsffile{ftxsad.eps}
\epsffile{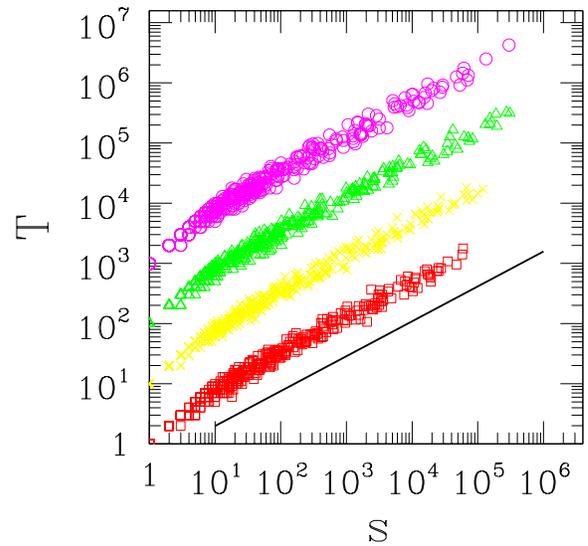}
\caption{Scatter plots of avalanche duration $T$ (number of lattice
sweeps) versus avalanche size (number of flipped sites), in the
adiabatic regime. Bottom to top: $L=40$, $80$, $160$, $320$; 
%$1,500$ events displayed for each lattice $L$.
$500$ events displayed for each lattice $L$.
Plots successively shifted upwards by a factor of 10 on vertical scale, to
avoid superposition. Straight line has slope $0.58$ (see text).}
\label{fig:txsad}
\end{center}
\end{figure}
In Figure~\ref{fig:txsad} one sees that in the adiabatic
regime, apart from the $L$--dependent
upper cutoff just mentioned, there is no distinguishable influence of
finite lattice width on the distribution of avalanche durations $T$
against size $s$. 

For small events with $s \lesssim 10$ flipping sites,
duration varies approximately linearly with size (collections of
independent, localized flippings). 
For larger, collective events, non-trivial scaling
takes hold; the relationship is described by
the power law $T \sim s^{\sigma\nu z}$ (notation borrowed from 
Refs.~\onlinecite{pds99,ks00}). Least-squares fits of data
excluding small avalanches up to $s_{min} = 10 - 20$ give
$\sigma\nu z = 0.58 \pm 0.01$, very similar to that
previously obtained in simulations of the  nucleation model,
variously quoted as $0.57 \pm 0.03$ (Ref.~\onlinecite{pds99})
or 0.58 (Ref.~\onlinecite{ks00}). 
Thus, in this aspect at least, there is universality between
single-interface and nucleation pictures.

%{\bf Referee's point (4)}

Similar values of $\sigma\nu z$ are obtained also
from slightly different variants of interface 
models~\cite{zap,zcds,dz99,dz00,lnst}.
In particular, simulations reported in Ref.~\onlinecite{lnst} give
$z=1.56 \pm 0.06$ and the roughness exponent $\zeta=0.75 \pm 0.02$,
which, together with the scaling relation $\sigma^{-1}=\nu(2+\zeta)$
(see Eqs. (34)--(37) of Ref.~\onlinecite{zcds}), yields
$\sigma\nu z=0.57 \pm 0.02$~.

%{\bf *****************************}

Furthermore, the distribution of durations fits reasonably well to a
power law with exponential tail, similarly to the size distribution
Eq.~(\ref{eq:psize}):
\begin{equation}
P(T) \propto T^{-\alpha}\exp\left(-T/T_0\right)\ \ ,
\label{eq:dur}
\end{equation}
where, from standard probability theory, $\alpha=1+ (\tau-1)/\sigma\nu z$
 ($=1.5 \pm 0.2$ from the values of $\tau$ and $\sigma\nu z$ quoted
above).
Indeed, analyzing the data shown in Fig.~\ref{fig:txsad} one gets $\alpha$
in the range $1.3-1.5$. The cutoff $T_0$ is expected to scale with the
size cutoff $s_0$ as $T_0 \sim s_0^{\sigma\nu z}$, therefore (with 
$s_0 \sim L^{1.4\pm 0.1}$~\cite{us}), one must have $T_0 \sim L^{0.81 \pm
0.06}$. Direct analysis of data gives the latter exponent varying in the
range $0.65-0.9$, broadly compatible with this.

Next, we gauge the effects of varying driving rates on the size-duration
relationship. For sufficiently fast driving rates,
one expects coalescence of avalanches which would be separate events in
the adiabatic regime. 

\begin{figure}
\epsfxsize=8.4cm
\begin{center}
\leavevmode
%\epsffile{ftxsvr.eps}
\epsffile{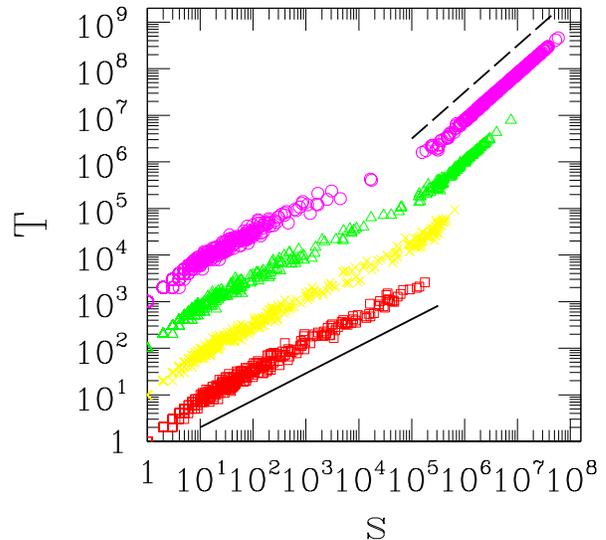}
\caption{Scatter plots of avalanche duration $T$ versus avalanche size, 
for $L=80$ and different driving rates $\Delta$. Bottom to top:
$\Delta=10^{-5}$, $5\times 10^{-5}$,  $7.5\times 10^{-5}$ and $10^{-4}$; 
%$1,500$ events displayed for first three, $3,500$ for $\Delta=10^{-4}$.
$500$ events displayed for first three, $1,000$ for $\Delta=10^{-4}$.
Plots successively shifted upwards by a factor of 10 on vertical scale, to
avoid superposition. Full straight line has slope $0.58$; dashed 
line has unitary slope (see text).
}
\label{fig:txsvr}
\end{center}
\end{figure}

We start by fixing $L=80$. $\Delta$, the external
field increase at the start of each lattice sweep during an
avalanche, is the driving rate . In Figure~\ref{fig:txsvr} one has
 $\Delta =10^{-5}$, $5\times 10^{-5}$, $7.5\times 10^{-5}$, and
$10^{-4}$. 

The plot for $\Delta=10^{-5}$ is identical, apart from small fluctuations,
to the corresponding one for $L=80$ in the adiabatic regime, shown in
Fig.~\ref{fig:txsad}. We have checked that the same happens for the 
in-between driving rates $10^{-m}$, $m=8$, 7, 6 (as expected).
Again, least-squares fits give $\sigma\nu z = 0.58 \pm 0.01$  not only
for the whole set of $\Delta =10^{-5}$ data, but also for the initial
parts (with $s \lesssim 10^5$) of those with faster driving rates. These
latter plots will be discussed further ahead. 

We conclude that, for $L=80$ and within the range of $\Delta=0-10^{-5}$, 
there is
no influence of the driving rate on the $T-s$ relationship, including
the maximum avalanche size, which remains at $s_{\rm max}\simeq 10^5$.  
So far, the results: (i) $\tau=1.3 \pm 0.1$; (ii) $\alpha= 1.5 \pm 0.2$
and (iii) independence of behavior on driving rate (at least, within
fairly well
defined windows of $\Delta$, for given $L$)
show that the model under consideration shares the same
universality class of the interface model discussed in 
Refs.~\onlinecite{zap,zcds,dz99,dz00}, when dipolar interactions are
neglected.

\section{rate-dependent behavior}
\label{secIV}

We now investigate the different picture appearing for
higher driving rates, already shown in Fig.~\ref{fig:txsvr}. The
scatter plot of duration against size for $\Delta=5\times 10^{-5}$ shows
traces of a different behavior for $s > s_{\rm max}$ as a higher slope 
develops at that point. As $\Delta$ is further increased,
it can be seen that there is coexistence between
non-trivial scaling for sizes $\lesssim$ the adiabatic cutoff, and
$T \sim s$ behavior (i.e. interface motion with a finite velocity)  for
larger avalanche sizes.

A qualitative explanation of the above results goes as follows.
With $H_f(i)$ and $H_i(i+1)$ being, respectively, external field at the
end of the $i$-th avalanche and at the start of the $i+1$-th, typically
the corresponding gaps
$\delta H_i^{ad} \equiv
H_i(i+1)-H_f(i)$ in the adiabatic regime are larger than, say, several
times $\Delta$ for $\Delta \leq 10^{-5}$; therefore few avalanches merge
for such driving rates.
For larger $\Delta$ more gaps are closed, and the distribution changes
significantly.

In order to gain a quantitative understanding of this, at the same time
checking for a possible $L$-dependence, we study the
probability distribution of $\delta H^{ad}$ for different lattice
widths. 

%{\bf Referee's point (5)}

\begin{figure}
\epsfxsize=8.4cm
\begin{center}
\leavevmode
\epsffile{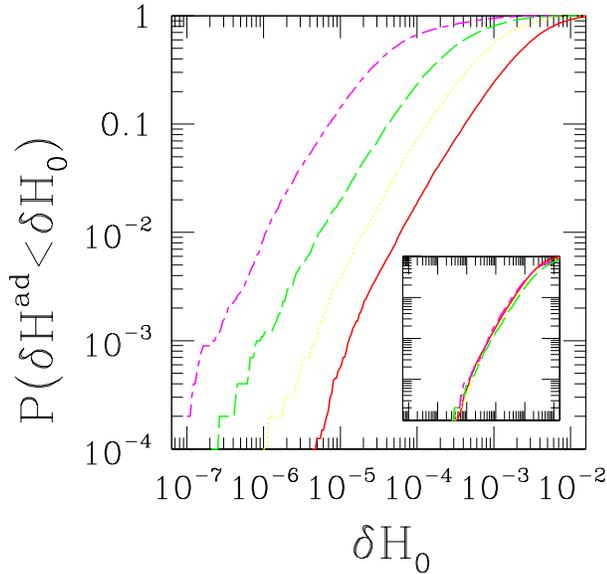}
\caption{Adiabatic regime: probability $P(\delta H^{ad}<\delta H_0)$ of
the gap between consecutive avalanches being smaller than $\delta H_0$, 
against $\delta H_0$. Left to right: $L=320$, 160, 80 and 40. 40,000
events for $L=40$, 10,000 for larger widths. Inset: $P(\delta
H^{ad}<\delta H_0)$ against $(L/40)^x\,\delta H_0$, $x=1.9$.
Same ranges as main plot, on both horizontal and vertical axes (see text).
}
\label{fig:dhip}
\end{center}
\end{figure}

In Fig.~\ref{fig:dhip} we show the cumulative probability
%
%{\bf *****************************}
%
$P(\delta H^{ad}<\delta H_0)$ of  $\delta H^{ad}$ being smaller than
$\delta H_0$. Before analyzing the $L-$ dependence of the curves, we
focus on $L=80$. 
One sees that $P(\delta H^{ad}<\delta H_0) \sim 3 \times 10^{-3},
7 \times 10^{-2}$, respectively for $\delta H_0=10^{-5}, 10^{-4}$. From 
Fig.~\ref{fig:txsvr}, the maximum avalanche duration is $T_{max} \sim
5 \times 10^3$, suggesting that the finite driving rate is irrelevant as
long as the quantity ${\cal P} \equiv {\cal P}(L,\Delta)
=T_{max}\,P(\delta H^{ad}<\Delta) \lesssim {\cal O}(10)$. On the contrary,
for $\Delta=10^{-4}$
Figs.~\ref{fig:txsad} and~\ref{fig:dhip}
show that ${\cal P} \sim {\cal O} (100)$. 

The inset of Fig.~\ref{fig:dhip} shows that $P(\delta H^{ad}<\delta H_0)$
is, to a good approximation, a function of $L^x\delta H_0$, where
the best collapse plots are obtained in the range $x=1.9 \pm 0.1$. 
Though at this point we are not able to advance an argument, it
may be that $x=2$ exactly.
It is thus clear that as $L$ increases
the range of $\Delta$ for which adiabatic
behavior dominates is shrunk. Recall that, on top of the scaling of
probabilities with $L^x\delta H_0$, $T_{max}$ (which scales with the
cutoff $T_0$) also depends on  $L^v$, $v \simeq 0.81$, see paragraph
below  Eq.~(\ref{eq:dur}). Thus, in the limit $L \to \infty$ one should
have adiabatic-like properties only strictly at $\Delta=0$. Before
attaching much significance to this, one must bear in mind the main
result of Ref.~\onlinecite{us}: in the present model, $L$ is closely
correlated with the (average, or typical) domain size in experimental
situations. Therefore, one does not have the usual concern in the study
of equilibrium second-order phase transitions, of trying to extract
the thermodynamic limit from finite-size scaling: finite-$L$ results
must be analyzed in their own right. Nevertheless, it is crucial to
investigate
the size dependence of relevant quantities, as done here, in order to
separate truly $L$-independent features from crossover behavior.

\begin{figure}
\epsfxsize=8.4cm
\begin{center}
\leavevmode
\epsffile{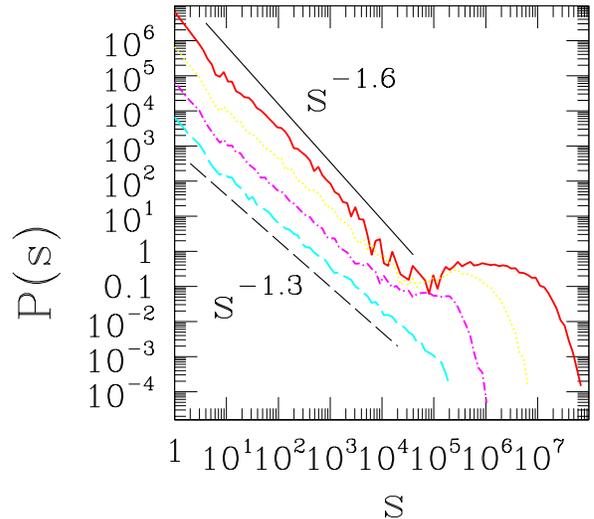}
\caption{Probability distributions of avalanche sizes for $L=80$.
Bottom to top: $\Delta=
10^{-5}$, $5 \times 10^{-5}$, $7.5 \times 10^{-5}$ and
$10^{-4}$. Curves successively shifted
upwards by a factor of 10 on vertical scale, to avoid superposition
(baseline is $\Delta=10^{-5}$ plot).
Straight lines have slopes
as indicated, and are intended as guides to the eye.
}
\label{fig:psnad}
\end{center}
\end{figure}

The departures from rate-independent behavior, as depicted 
in Fig.~\ref{fig:txsvr}, require further analysis.   
 A least-squares fit of data in that Figure for $\Delta=10^{-4}$, $10 
\leq s \leq 5
\times 10^4$ (the horizontal extent of the full line shown there)
gives $\sigma\nu z=0.58 \pm 0.01$. The events in this range of $s$ are 
$\sim 20$\% of the total number of avalanches; another 40\% are small
ones with $s< 10$, with the remaining 40\% larger than $5  \times 10^4$
sites. 
The overall probability density $P(s)$ for $\Delta=10^{-4}$, with
the customary logarithmic binning,
is displayed in Fig.~\ref{fig:psnad}, for the four driving rates
of Fig.~\ref{fig:txsvr}~.
%(as already mentioned when discussing duration versus
%size plots, the curves for all $\Delta \leq
%10^{-5}$ are identical, apart from small fluctuations). 
The full
straight line suggests that the  $\Delta=10^{-4}$ data with $1 \leq s
\lesssim 5 \times10^4$ can be fitted by a power law $P(s)\propto s^{-1.6}$.
 This larger effective value of $\tau$ can 
be understood by observing the depletion of events with sizes 
$10^4-10^5$ in the cross-over region of Fig.~\ref{fig:txsvr}.
Curves for $\Delta=5 \times 10^{-5}$ and $7.5 \times 10^{-5}$ display
intermediate behavior, with only an incipient
shoulder at $10^4 \lesssim s \lesssim 10^5$ for the former.
%{\bf Referee's point (2)}

We note that in earlier studies, both experimental~\cite{bdm94} and
numerical~\cite{bt96}, it was found that the exponents $\tau$ of
Eq.~(\ref{eq:psize}) and $\alpha$ of Eq.~(\ref{eq:dur}) {\em decrease}
as driving rate increases, apparently in contradiction with the above.
Also, a small shoulder can be seen in the data for alloys under stress
in Figure 1(a) of Ref.~\onlinecite{dz00}. We comment on each of these in
turn. 

Firstly, recent work~\cite{dz00} shows that the FeSi alloys of 
Ref.~\onlinecite{bdm94} and, e.g., amorphous alloys under
stress belong to different universality classes; while the former 
follow a mean-field description (giving rise to actual rate-dependent
exponents), the latter display rate-independent exponents
with values close to those found both for the
present model, and for the model of
Refs.~\onlinecite{zap,zcds,dz99,dz00} when dipolar interactions are
neglected (see remarks at the end of Sec.~\ref{std})~.

As regards the simulations of  Ref.~\onlinecite{bt96},
it may be that the model used here 
does not belong to the same dynamical
universality class as the (two-dimensional) random-field model with
vacancies used there. However, examination of the data
in Fig.~\ref{fig:psnad} suggests an alternative explanation, as follows.
Though fits of the straight-line parts of the distributions give
$\tau$ increasing with $\Delta$, fits of the whole sets of
data to a form such as Eq.~(\ref{eq:psize}) , with {\em two} free
parameters ($\tau$ and $s_0$) in fact give {\em decreasing} values
of $\tau$ for larger $\Delta$, on account of the large-$s$ shoulders .
Of course, this happens at the expense
of the quality of the fit; however, for less dramatic departures than
those shown in the Figure, the loss of quality might not be obvious.

Finally, the incipient shoulders shown in Ref.~\onlinecite{dz00} appear to
be {\it bona fide} candidates for the above description, as they
correspond to materials to which, so far, the present model seems
to fit well. We believe that a reexamination of experimental and
simulational raw data, in search of a coexistence of regimes, would be
worthwhile. As pointed out in Ref.~\onlinecite{bt96}, ``[a faster driving
rate] overdrives weaker pinning centers thus rendering the occurrence of
larger avalanches more probable''. For the present
model this indeed happens, only it does so at the expense of depleting   
the histograms of occurrence of events $\lesssim$ the respective maximum
size for the adiabatic regime. We expect the present study to motivate
further experimental and numerical investigation along these lines. 

%{\bf *****************************}

\section{intermittency}
\label{secV}

\begin{figure}
\epsfxsize=8.4cm
\begin{center}
\leavevmode
\epsffile{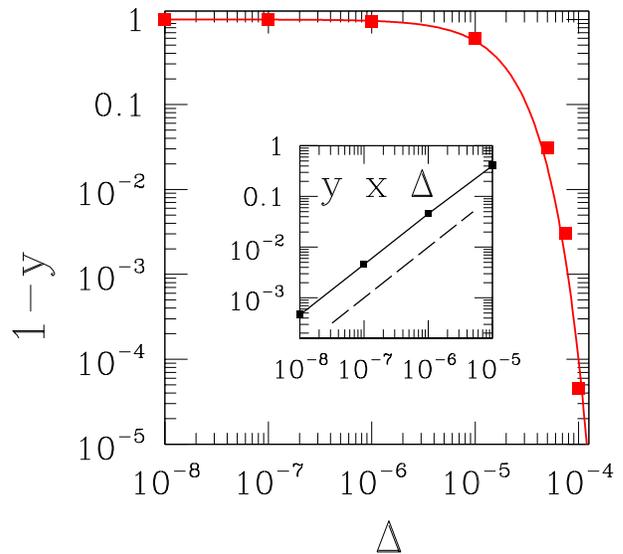}
\caption{Plot of $1-y$ ($y$ is the fraction of the total duration spent
during avalanches)
against driving rate $\Delta$, for $L=80$. Solid curve is fit to
$1-y=\exp[-(\Delta/\Delta_0)^a]$, $a=1.2$, $\Delta_0=1.57 \times 10^{-5}$.
Inset: $y$ against $\Delta$ for slow driving; dashed line has unitary 
slope. 
}
\label{fig:y}
\end{center}
\end{figure}

It is also important to analyze how the intermittency of events is
gradually lost as more and more avalanches coalesce with increasing
driving
rate.  This can be done by defining $y$ as the fraction of time spent
during avalanches. The duration of an avalanche being given by $T(i) =
(H_f(i)-H_i(i))/\Delta$, the overall duration of a simulation with $N$
events is   $T_N=(H_f(N)-H_i(1))/\Delta$, therefore $y=(1/T_N)\sum_{i=1}^N
T(i)=[\sum_{i=1}^N(H_f(i)-H_i(i)]/[H_f(N)-H_i(1)]$.
In the adiabatic regime $\Delta=0$, avalanches are instantaneous, and $y=0$.
As $\Delta$ is increased, avalanches will be observed part of the time, 
so $0<y<1$,  and in the limit 
of large $\Delta$, one expects that a depinning transition will lead to $y=1$.
Figure ~\ref{fig:y} shows the plot of $1-y$ versus $\Delta$ for $L=80$. 
The best non-linear fit of the whole set of 
points is given by $1-y=\exp[-(\Delta/\Delta_0)^a]$ with the optimum
values of the free parameters: $a=1.2$  and $\Delta_0=1.57\times10^{-5}$. 
The inset shows the rate-independent regime corresponding to $y\ll 1$,
where one clearly has $y\sim \Delta$. Such linearity is to be expected,
as essentially the same events occur for any $\Delta$ in this interval:
for a given avalanche, $H_f(i)-H_i(i)$ is proportional to $\Delta$, while
the denominator $H_f(N)-H_i(1)$ is unchanged. The fact that the
best-fitting value of $a$ is 1.2 indicates that the decay of $1-y$ for
large $\Delta$
is in fact faster than $\exp[-(\Delta/\Delta_0)^{1.2}]$, so one is having
a compensation among the different regions of the plot, in order to
minimize the overall deviation. 
A fit of the subset of data for $\Delta \geq 10^{-5}$ indeed gives $a
\simeq 2$. 

\section{power spectra}
\label{secVI}
Power spectra and their analysis are an important tool for the
understanding of noise in disordered systems; specifically for
the Barkhausen effect, see e.g. 
Ref.~\onlinecite{ks00} for a recent survey of 
results and references; see also Ref.~\onlinecite{dz01}. 

We shall always make $L=80$ in this section. With the unit of time
given by a lattice sweep, the maximum frequency to be analyzed is
$\omega_{M}=1/2$, as standard Fourier theory prescribes.

We attempt to concentrate on the non-trivial scaling regime. To this
end, we recall from Fig.~\ref{fig:txsvr} that, whatever the driving rate,
events of size $s \lesssim 10$ sites (of which there is always a large
fraction of the total) are collections of independent,
localized flippings. Thus their duration is proportional to size.
Therefore, when looking at power-spectrum data with frequencies
$\omega \gtrsim 0.1$, one will have a strong input from such
non-critical flippings. Though one might think of ways to expurgate the
respective contributions, we shall simply restrict ourselves to
the frequency region $\omega \leq 0.1$. As shown in the following,
this still leaves a suitably wide window of observation, in what we
call intermediate-frequency range (for $\omega \to 0$ the power spectrum
goes flat, as details of the temporal  series are washed out on long time
scales).

The discussion of power spectra revolves mainly around the power $t$ with
which decay  sets in, in the intermediate-frequency range: $P(\omega) \sim
\omega^{-t}$.
The following points were made in Ref.~\onlinecite{ks00}:
(1)  While one has $t=2$ for white
noise and mean-field descriptions, and early studies of the Barkhausen
effect predicted 
$t=(3-\tau)/\sigma\nu z$ ($=2.9 \pm 0.2$ for the present model, 
with $\tau=1.3 \pm 0.1$, $\sigma\nu z = 0.58 \pm 0.01$), it
was found that for $\tau < 2$ one should have $t=1/\sigma\nu z$
 ($=1.72\pm 0.03$ here), instead of $(3-\tau)/\sigma\nu z$. Several
experimental and simulational results were shown to be compatible
with the latter finding. (2) It was remarked that the result of earlier
calculations of the power spectrum~\cite{jcf} for sandpile models, which
gave $P(\omega) \sim  \omega^{-2}$ and were cited as an explanation for
such behavior in (among other systems) Barkhausen noise, depends
crucially on the assumption that the avalanche shape can be approximated
by a box function: $V(t)=S/T$ ($0<t<T$) for an avalanche of size $S$ and
duration $T$.

In the present model, one can tune the degree to which the internal
structure of an avalanche is taken into account, by varying $\Delta$.
While for $\Delta=0$ all events are seen as spikes of zero duration, the
internal fluctuations within avalanches become much more noticeable
as $\Delta$ increases, even still within the adiabatic regime. Recall
that, for $\Delta=10^{-5}$, $L=80$, avalanches take up $\sim 40\%$ of
the total duration of a simulation, see inset of Fig.~\ref{fig:y}.

\begin{figure}
\epsfxsize=8.4cm
\begin{center}
\leavevmode
%\epsffile{fps.eps}
\epsffile{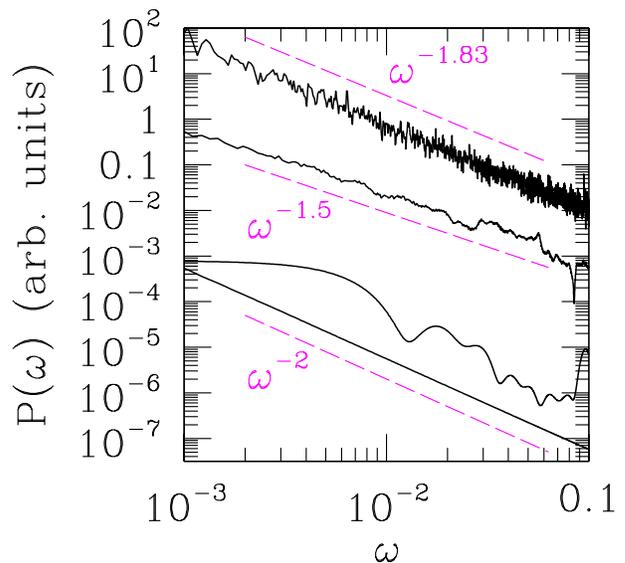}
\caption{Continuous lines: power spectra for $L=80$ and various driving
rates. Bottom to
top, $\Delta=10^{-7}$ $10^{-6}$, $10^{-5}$ and $10^{-4}$. The vertical 
coordinates have been shifted by differing amounts for each plot, so
they could fit in a single graph. Dashed lines have slopes as indicated,
and are intended as guides to the eye.
}
\label{fig:ps}
\end{center}
\end{figure}

Below, we set out to probe points (1) and (2). Accordingly, in 
Fig.~\ref{fig:ps} we plot
the power spectra for $\Delta=10^{-7}$ (deep within the adiabatic regime),
$10^{-6}$, $10^{-5}$ and $10^{-4}$.  

Starting from the slowest driving rate considered, one sees that
disregarding the internal structure of avalanches indeed yields 
a dependence of the power spectrum on $\omega^{-2}$. This is entirely
consistent with point (2) above.
The next graph, $\Delta=10^{-6}$, is somewhat difficult to interpret on
its own. However, the trend becomes clearer when the $\Delta=10^{-5}$ data
are taken into account: as more details of the intra-avalanche
fluctuations enter into the spectrum, its decay becomes slower, $\sim 
\omega^{-1.5}$.
Though the numerical values are somewhat off the mark, one definitely sees
that the trend is towards $t=1/\sigma\nu z$ when intra-avalanche
correlations are considered, as opposed to the alternative
$t=(3-\tau)/\sigma\nu z$. This is in support of point (1) above, showing
that very likely  the present model behaves similarly to the random-field
one of Ref.~\onlinecite{ks00}, in this respect. 
    
Finally, for $\Delta=10^{-4}$ there is an apparent trend reversal;
from a least-squares fit of data in the range of $\omega$ shown, one
has the exponent $1.83 \pm 0.01$. It must be recalled that one is
then clearly in the coexistence
regime explained above (that is, away from a purely
intermittent, adiabatic framework); therefore there is strong influence
of inter-avalanche correlations. It is then not surprising that the
picture starts to differ e.g. from that of  Ref.~\onlinecite{ks00},
where only inter-avalanche correlations were taken into account.  

Clearly, more work is needed to sort out this latter point. 
As one goes deeper into the depinned regime, it may well be that
the power spectrum decay returns to the $\omega^{-2}$ form
characteristic of uncorrelated events. In such a scenario, the above
value of $t=1.83$ would in fact be an effective exponent, marking a
crossover towards $t=2$.   

\section{conclusions}
\label{conc}
We have used a single-interface model with an adjustable finite driving 
rate to simulate
the time sequence of Barkhausen jumps of a three-dimensional system.
Direct observation of duration as a function of size determines 
the existence of 
two dynamical regimes regarding the $\sigma\nu z$ exponent, that is, for low 
driving rates ($\Delta<\Delta_c(L)$), $\sigma\nu z=0.58$ up to a 
limiting event size, 
and the avalanche dynamics is basically rate independent. For higher driving 
rates ($\Delta>\Delta_c(L)$)
the previous regime coexists with one for which $\sigma\nu z=1$. 
The rate dependency 
of the second regime is evident as we analyze the probability 
distribution of avalanches 
sizes: for high driving rates it deviates considerably from the usual form, 
$P(s)\propto s^{-\tau} \exp(-s/s_0)$. The passage from one regime to 
the other is rather sharp, 
and the corresponding value of driving rate, $\Delta_c$, depends on the 
system size: an infinite system will have a rate-independent dynamics for 
$\Delta=0$ only, that is, $\Delta_c(\infty)=0$.

Considering only the power-law portion of 
the $P(s)$ graphs for finite driving rate, we find exponents that increase with 
the driving rate, in apparent contradiction with previous 
theoretical and experimental results.
On the other hand, when the fitting assumes a power law with an
exponential cut-off, and the whole set of data is taken into account,
the effective value of $\tau$ decreases as the driving rate is increased. 
It is clear though
that this form does not provide an adequate description of  
the simulated data for $\Delta>\Delta_c$, so we believe that 
this may be at least part of the explanation for inconsistency 
in previously reported values of $\tau$.

The power spectra for various driving rates clearly show that 
with increasing driving rates, intra-avalanche correlations become 
more relevant, as 
the time scale involved
reveals details  of events occuring with a finite duration. 
A direct consequence is the relation
$P(\omega) \sim \omega^{-t}$ observed: as the internal structure of the
avalanches is probed, $t$
decreases from the value $t=2$, characteristic of adiabatic time series. 

In summary, we have studied the effect of a finite driving rate 
in the scaling properties of the Barkhausen noise. As our 
ultimate goal is the description of experimental results, it is 
important to understand the limitations involved in real experiments 
as compared to computational ones. In principle simulations may use 
any value for the driving rate, as well as any system size, or 
at least we may say that our choice of values is broad as compared to 
real experiments. A typical experiment usually has driving rates spanning
only one order of magnitude, and  values 
of domain sizes predetermined by the fabrication conditions of the 
sample and applied stress. So, as the experiment is designed 
with these parameters, the scaling regime is basically already chosen. 
Any further comparison with theoretical or simulation results 
must be careful in the sense that the same regime has to be studied. 

\acknowledgments
The authors thank Belita Koiller and Mark Robbins for interesting
discussions,
and Brazilian agencies CNPq (grants \# 30.1692/81.5 and 30.1057/90.7),
FAPERJ (grants \# E26--171.447/97 and \# E26--151.869/2000) and FUJB-UFRJ 
for financial support. We thank a
referee for interesting comments, and for drawing our attention to
relevant references.

%% BIBLIOGRAPHY

\end{document}